\documentclass[11pt,oneside,american,reqno]{amsart}
\usepackage[T1]{fontenc}
\usepackage{xcolor}
\usepackage{pdfcolmk}
\usepackage{verbatim}
\usepackage{amsbsy}
\usepackage{amstext}
\usepackage{amsthm}
\usepackage{amssymb}
\usepackage{undertilde}
\usepackage{xargs}[2008/03/08]
\PassOptionsToPackage{normalem}{ulem}
\usepackage{ulem}

\makeatletter

\providecolor{lyxadded}{rgb}{0,0,1}
\providecolor{lyxdeleted}{rgb}{1,0,0}

\DeclareRobustCommand{\lyxsout}[1]{\ifx\\#1\else\sout{#1}\fi}

\numberwithin{equation}{section}
\theoremstyle{plain}
\newtheorem{thm}{\protect\theoremname}
  \theoremstyle{remark}
  \newtheorem{rem}[thm]{\protect\remarkname}
  \theoremstyle{definition}
  \newtheorem{example}[thm]{\protect\examplename}
  \theoremstyle{remark}
  \newtheorem*{acknowledgement*}{\protect\acknowledgementname}

\usepackage{pdfsync} 
\usepackage[all]{xy}
\newcommand{\ie}{\textit{i.e.}}
\newcommand{\eg}{\textit{e.g.}}
\usepackage[lining,tabular]{fbb} 
\usepackage[libertine,bigdelims]{newtxmath}
\usepackage[cal=boondoxo,bb=boondox,frak=boondox]{mathalfa}

   \usepackage{bm}
\usepackage{bm}
\renewcommand{\mathbf}{\bm}
 \theoremstyle{plain}
  
\makeatother

\usepackage{babel}
  \providecommand{\acknowledgementname}{Acknowledgement}
  \providecommand{\examplename}{Example}
  \providecommand{\remarkname}{Remark}
\providecommand{\theoremname}{Theorem}

\begin{document}

\global\long\def\ga{\alpha}
\global\long\def\gb{\beta}
\global\long\def\ggm{\gamma}
\global\long\def\go{\omega}
\global\long\def\gs{\sigma}
\global\long\def\gd{\delta}
\global\long\def\gD{\Delta}
\global\long\def\vph{\varphi}
\global\long\def\gf{\varphi}
\global\long\def\gk{\kappa}
\global\long\def\gl{\lambda}
\global\long\def\gz{\zeta}
\global\long\def\gh{\eta}
\global\long\def\gy{\upsilon}

\global\long\def\eps{\varepsilon}
\global\long\def\epss#1#2{\varepsilon_{#2}^{#1}}
\global\long\def\ep#1{\eps_{#1}}

\global\long\def\wh#1{\widehat{#1}}

\global\long\def\spec#1{\textsf{#1}}

\global\long\def\ui{\wh{\boldsymbol{\imath}}}
\global\long\def\uj{\wh{\boldsymbol{\jmath}}}
\global\long\def\uk{\widehat{\boldsymbol{k}}}

\global\long\def\uI{\widehat{\mathbf{I}}}
\global\long\def\uJ{\widehat{\mathbf{J}}}
\global\long\def\uK{\widehat{\mathbf{K}}}

\global\long\def\bs#1{\boldsymbol{#1}}
\global\long\def\vect#1{\mathbf{#1}}
\global\long\def\bi#1{\textbf{\emph{#1}}}

\global\long\def\uv#1{\widehat{\boldsymbol{#1}}}
\global\long\def\cross{\times}

\global\long\def\ddt{\frac{\dee}{\dee t}}
\global\long\def\dbyd#1{\frac{\dee}{\dee#1}}
\global\long\def\dby#1#2{\frac{\partial#1}{\partial#2}}

\global\long\def\vct#1{\mathbf{#1}}

\global\long\def\partialby#1#2{\frac{\partial#1}{\partial x^{#2}}}
\newcommandx\parder[2][usedefault, addprefix=\global, 1=]{\frac{\partial#2}{\partial#1}}

\global\long\def\reals{\mathbb{R}}

\global\long\def\rthree{\reals^{3}}
\global\long\def\rsix{\reals^{6}}
\global\long\def\rn{\reals^{n}}
\global\long\def\rt#1{\reals^{#1}}

\global\long\def\les{\leqslant}
\global\long\def\ges{\geqslant}

\global\long\def\dee{\textrm{d}}
\global\long\def\di{d}

\global\long\def\from{\colon}
\global\long\def\tto{\longrightarrow}
\global\long\def\lmt{\longmapsto}

\global\long\def\abs#1{\left|#1\right|}

\global\long\def\isom{\cong}

\global\long\def\comp{\circ}

\global\long\def\cl#1{\overline{#1}}

\global\long\def\fun{\varphi}

\global\long\def\interior{\textrm{Int}\,}

\global\long\def\sign{\textrm{sign}\,}
\global\long\def\sgn#1{(-1)^{#1}}
\global\long\def\sgnp#1{(-1)^{\abs{#1}}}

\global\long\def\dimension{\textrm{dim}\,}

\global\long\def\esssup{\textrm{ess}\,\sup}

\global\long\def\ess{\textrm{{ess}}}

\global\long\def\kernel{\mathop{\textrm{Kernel}}}

\global\long\def\support{\textrm{supp}\,}

\global\long\def\image{\textrm{Image}\,}

\global\long\def\diver{\mathop{\textrm{div}}}

\global\long\def\sp{\mathop{\textrm{span}}}

\global\long\def\resto#1{|_{#1}}
\global\long\def\incl{\iota}
\global\long\def\iden{\imath}
\global\long\def\idnt{\textrm{Id}}
\global\long\def\rest{\rho}
\global\long\def\extnd{e_{0}}

\global\long\def\proj{\textrm{pr}}

\global\long\def\L#1{L\bigl(#1\bigr)}
\global\long\def\LS#1{L_{S}\bigl(#1\bigr)}

\global\long\def\ino#1{\int_{#1}}

\global\long\def\half{\frac{1}{2}}
\global\long\def\shalf{{\scriptstyle \half}}
\global\long\def\third{\frac{1}{3}}

\global\long\def\empt{\varnothing}

\global\long\def\paren#1{\left(#1\right)}
\global\long\def\bigp#1{\bigl(#1\bigr)}
\global\long\def\biggp#1{\biggl(#1\biggr)}
\global\long\def\Bigp#1{\Bigl(#1\Bigr)}

\global\long\def\braces#1{\left\{  #1\right\}  }
\global\long\def\sqbr#1{\left[#1\right]}
\global\long\def\anglep#1{\left\langle #1\right\rangle }

\global\long\def\lsum{{\textstyle \sum}}

\global\long\def\bigabs#1{\bigl|#1\bigr|}

\global\long\def\vs{\textbf{W}}
\global\long\def\avs{\textbf{V}}
\global\long\def\affsp{\mathbf{A}}
\global\long\def\pt{p}

\global\long\def\vbase{e}
\global\long\def\sbase{\mathbf{e}}
\global\long\def\msbase{\mathfrak{e}}
\global\long\def\vect{v}
\global\long\def\dbase{\sbase}

\global\long\def\stp{\text{\small\ensuremath{\bigodot}}}
\global\long\def\tp{\text{\small\ensuremath{\bigotimes}}}

\global\long\def\mi#1{\boldsymbol{#1}}
\global\long\def\mii{\mi I}
\global\long\def\mie#1#2{#1_{1}\dots#1_{#2}}
\global\long\def\asmi#1{#1}

\global\long\def\miem#1#2#3{#1_{#3}\cdots#1_{#2}}
\global\long\def\ordr#1{\left\langle #1\right\rangle }

\global\long\def\symm#1{\paren{#1}}
\global\long\def\smtr{\mathcal{S}}

\global\long\def\perm{p}
\global\long\def\sperm{\mathcal{P}}

\global\long\def\fall{,\quad\text{for all}\quad}

\global\long\def\inosum{,\qquad\text{no sum on }\mii}
\global\long\def\nosumg#1{,\qquad\text{no sum on }#1}

\global\long\def\commen#1{\qquad\text{#1}}

\global\long\def\oneto{1,\dots,}

\global\long\def\lisub#1#2#3{#1_{1}#2\dots#2#1_{#3}}

\global\long\def\lisup#1#2#3{#1^{1}#2\dots#2#1^{#3}}

\global\long\def\lisubb#1#2#3#4{#1_{#2}#3\dots#3#1_{#4}}

\global\long\def\lisubbc#1#2#3#4{#1_{#2}#3\cdots#3#1_{#4}}

\global\long\def\lisubbwout#1#2#3#4#5{#1_{#2}#3\dots#3\widehat{#1}_{#5}#3\dots#3#1_{#4}}

\global\long\def\lisubc#1#2#3{#1_{1}#2\cdots#2#1_{#3}}

\global\long\def\lisupc#1#2#3{#1^{1}#2\cdots#2#1^{#3}}

\global\long\def\lisupp#1#2#3#4{#1^{#2}#3\dots#3#1^{#4}}

\global\long\def\lisuppc#1#2#3#4{#1^{#2}#3\cdots#3#1^{#4}}

\global\long\def\lisuppwout#1#2#3#4#5#6{#1^{#2}#3#4#3\wh{#1^{#6}}#3#4#3#1^{#5}}

\global\long\def\lisubbwout#1#2#3#4#5#6{#1_{#2}#3#4#3\wh{#1}_{#6}#3#4#3#1_{#5}}

\global\long\def\lisubwout#1#2#3#4{#1_{1}#2\dots#2\widehat{#1}_{#4}#2\dots#2#1_{#3}}

\global\long\def\lisupwout#1#2#3#4{#1^{1}#2\dots#2\widehat{#1^{#4}}#2\dots#2#1^{#3}}

\global\long\def\lisubwoutc#1#2#3#4{#1_{1}#2\cdots#2\widehat{#1}_{#4}#2\cdots#2#1_{#3}}

\global\long\def\twp#1#2#3{\dee#1^{#2}\wedge\dee#1^{#3}}

\global\long\def\thp#1#2#3#4{\dee#1^{#2}\wedge\dee#1^{#3}\wedge\dee#1^{#4}}

\global\long\def\fop#1#2#3#4#5{\dee#1^{#2}\wedge\dee#1^{#3}\wedge\dee#1^{#4}\wedge\dee#1^{#5}}

\global\long\def\idots#1{#1\dots#1}
\global\long\def\icdots#1{#1\cdots#1}

\global\long\def\norm#1{\|#1\|}

\global\long\def\nonh{\heartsuit}

\global\long\def\nhn#1{\norm{#1}^{\nonh}}

\global\long\def\trps{^{{\scriptscriptstyle \textsf{T}}}}

\global\long\def\testfuns{\mathcal{D}}

\global\long\def\ntil#1{\tilde{#1}{}}

\global\long\def\eucl{E}

\global\long\def\mind{\alpha}
\global\long\def\vb{W}
\global\long\def\vbp{\pi}

\global\long\def\man{\mathcal{M}}
\global\long\def\odman{\mathcal{N}}
\global\long\def\subman{\mathcal{A}}

\global\long\def\vbt{\mathcal{E}}
\global\long\def\fib{\mathbf{V}}
\global\long\def\vbts{W}
\global\long\def\avb{U}

\global\long\def\chart{\varphi}
\global\long\def\vbchart{\Phi}

\global\long\def\jetb#1{J^{#1}}
\global\long\def\jet#1{j^{1}(#1)}
\global\long\def\tjet{\tilde{\jmath}}

\global\long\def\Jet#1{J^{1}(#1)}

\global\long\def\jetm#1{j_{#1}}

\global\long\def\coj{\mathfrak{d}}

\global\long\def\alt{\mathfrak{A}}

\global\long\def\pou{\eta}

\global\long\def\ext{{\textstyle \bigwedge}}
\global\long\def\forms{\Omega}

\global\long\def\dotwedge{\dot{\mbox{\ensuremath{\wedge}}}}

\global\long\def\vel{\theta}

\global\long\def\contr{\raisebox{0.4pt}{\mbox{\ensuremath{\lrcorner}}}\,}
\global\long\def\fcontr{\raisebox{0.4pt}{\mbox{\ensuremath{\llcorner}}}\,}

\global\long\def\lie{\mathcal{L}}

\global\long\def\ssym#1#2{\ext^{#1}T^{*}#2}

\global\long\def\sh{^{\sharp}}

\global\long\def\spc{\mathcal{S}}
\global\long\def\sptm{\mathcal{E}}
\global\long\def\evnt{e}
\global\long\def\frame{\Phi}

\global\long\def\timeman{\mathcal{T}}
\global\long\def\zman{t}
\global\long\def\dims{n}
\global\long\def\m{\dims-1}
\global\long\def\dimw{m}

\global\long\def\wc{z}

\global\long\def\fourv#1{\mbox{\ensuremath{\mathfrak{#1}}}}

\global\long\def\pbform#1{\utilde{#1}}
\global\long\def\util#1{\raisebox{-5pt}{\ensuremath{{\scriptscriptstyle \sim}}}\!\!\!#1}

\global\long\def\utilJ{\util J}

\global\long\def\utilRho{\util{\rho}}

\global\long\def\body{B}
\global\long\def\man{\mathcal{M}}
\global\long\def\var{\mathcal{V}}
\global\long\def\base{\mathcal{X}}
\global\long\def\fb{\mathcal{Y}}
\global\long\def\dimb{n}
\global\long\def\dimf{m}

\global\long\def\bdry{\partial}

\global\long\def\gO{\varOmega}

\global\long\def\reg{\mathcal{R}}
\global\long\def\bdrr{\bdry\reg}

\global\long\def\bdom{\bdry\gO}

\global\long\def\bndo{\partial\gO}

\global\long\def\pis{x}
\global\long\def\xo{\pis_{0}}

\global\long\def\pib{X}

\global\long\def\pbndo{\Gamma}
\global\long\def\bndoo{\pbndo_{0}}
 \global\long\def\bndot{\pbndo_{t}}

\global\long\def\cloo{\cl{\gO}}

\global\long\def\nor{\mathbf{n}}
\global\long\def\Nor{\mathbf{N}}

\global\long\def\dA{\,\dee A}

\global\long\def\dV{\,\dee V}

\global\long\def\eps{\varepsilon}

\global\long\def\vf{w}

\global\long\def\avf{u}

\global\long\def\stn{\varepsilon}

\global\long\def\rig{r}

\global\long\def\rigs{\mathcal{R}}

\global\long\def\qrigs{\!/\!\rigs}

\global\long\def\qd{\!/\,\!\kernel\diffop}

\global\long\def\dis{\chi}
\global\long\def\conf{\kappa}
\global\long\def\csp{\mathcal{Q}}

\global\long\def\embds{\textrm{Emb}}

\global\long\def\fc{F}

\global\long\def\st{\sigma}

\global\long\def\bfc{\mathbf{b}}

\global\long\def\sfc{\mathbf{t}}

\global\long\def\stm{\varsigma}
\global\long\def\std{S}
\global\long\def\tst{\sigma}

\global\long\def\nhs{Y}

\global\long\def\soc{Z}

\global\long\def\tran{\mathrm{tr}}

\global\long\def\slf{R}

\global\long\def\sts{\varSigma}
\global\long\def\spstd{\mathfrak{S}}
\global\long\def\sptst{\mathfrak{T}}

\global\long\def\spsb{\text{\Large\ensuremath{\Delta}}}

\global\long\def\ebdfc{T}
\global\long\def\optimum{\st^{\textrm{opt}}}
\global\long\def\scf{K}

\global\long\def\pform{\varsigma}
\global\long\def\vform{\beta}
\global\long\def\sform{\tau}
\global\long\def\flow{J}
\global\long\def\n{\m}
\global\long\def\cmap{\mathfrak{t}}
\global\long\def\vcmap{\varSigma}

\global\long\def\mvec{\mathfrak{v}}
\global\long\def\mveco#1{\mathfrak{#1}}
\global\long\def\smbase{\mathfrak{e}}
\global\long\def\spx{\simp}

\global\long\def\hp{H}
\global\long\def\ohp{h}

\global\long\def\hps{G_{\dims-1}(T\spc)}
\global\long\def\ohps{G_{\dims-1}^{\perp}(T\spc)}
\global\long\def\hpsx{G_{\dims-1}(\tspc)}
\global\long\def\ohpsx{G_{\dims-1}^{\perp}(\tspc)}

\global\long\def\fbun{F}

\global\long\def\flowm{\Phi}

\global\long\def\tgb{T\spc}
\global\long\def\ctgb{T^{*}\spc}
\global\long\def\tspc{T_{\pis}\spc}
\global\long\def\dspc{T_{\pis}^{*}\spc}

\global\long\def\fflow{\fourv J}
\global\long\def\fvform{\mathfrak{b}}
\global\long\def\fsform{\mathfrak{t}}
\global\long\def\fpform{\mathfrak{s}}

\global\long\def\maxw{\mathfrak{g}}
\global\long\def\frdy{\mathfrak{f}}
\global\long\def\ptnl{A}

\global\long\def\sobp#1#2{W_{#2}^{#1}}

\global\long\def\inner#1#2{\left\langle #1,#2\right\rangle }

\global\long\def\fields{\sobp pk(\vb)}

\global\long\def\bodyfields{\sobp p{k_{\partial}}(\vb)}

\global\long\def\forces{\sobp pk(\vb)^{*}}

\global\long\def\bfields{\sobp p{k_{\partial}}(\vb\resto{\bndo})}

\global\long\def\loadp{(\sfc,\bfc)}

\global\long\def\strains{\lp p(\jetb k(\vb))}

\global\long\def\stresses{\lp{p'}(\jetb k(\vb)^{*})}

\global\long\def\diffop{D}

\global\long\def\strainm{E}

\global\long\def\incomps{\vbts_{\yieldf}}

\global\long\def\devs{L^{p'}(\eta_{1}^{*})}

\global\long\def\incompsns{L^{p}(\eta_{1})}

\global\long\def\testf{\mathcal{D}}
\global\long\def\dists{\mathcal{D}'}

\global\long\def\codiv{\boldsymbol{\partial}}

\global\long\def\currof#1{\tilde{#1}}

\global\long\def\chn{c}
\global\long\def\chnsp{\mathbf{F}}

\global\long\def\current{T}
\global\long\def\curr{R}

\global\long\def\gdiv{\bdry\textrm{iv\,}}

\global\long\def\prop{P}

\global\long\def\aprop{Q}

\global\long\def\flux{\omega}
\global\long\def\aflux{S}

\global\long\def\fform{\tau}

\global\long\def\dimn{n}

\global\long\def\sdim{{\dimn-1}}

\global\long\def\contrf{{\scriptstyle \smallfrown}}

\global\long\def\prodf{{\scriptstyle \smallsmile}}

\global\long\def\ptnl{\varphi}

\global\long\def\form{\omega}

\global\long\def\dens{\rho}

\global\long\def\simp{s}
\global\long\def\ssimp{\Delta}
\global\long\def\cpx{K}

\global\long\def\cell{C}

\global\long\def\chain{B}

\global\long\def\ach{A}

\global\long\def\coch{X}

\global\long\def\scale{s}

\global\long\def\fnorm#1{\norm{#1}^{\flat}}

\global\long\def\chains{\mathcal{A}}

\global\long\def\ivs{\boldsymbol{U}}

\global\long\def\mvs{\boldsymbol{V}}

\global\long\def\cvs{\boldsymbol{W}}

\global\long\def\cee#1{C^{#1}}

\global\long\def\lone{L^{1}}

\global\long\def\linf{L^{\infty}}

\global\long\def\lp#1{L^{#1}}

\global\long\def\ofbdo{(\bndo)}

\global\long\def\ofclo{(\cloo)}

\global\long\def\vono{(\gO,\rthree)}

\global\long\def\vonbdo{(\bndo,\rthree)}
\global\long\def\vonbdoo{(\bndoo,\rthree)}
\global\long\def\vonbdot{(\bndot,\rthree)}

\global\long\def\vonclo{(\cl{\gO},\rthree)}

\global\long\def\strono{(\gO,\reals^{6})}

\global\long\def\sob{W_{1}^{1}}

\global\long\def\sobb{\sob(\gO,\rthree)}

\global\long\def\lob{\lone(\gO,\rthree)}

\global\long\def\lib{\linf(\gO,\reals^{12})}

\global\long\def\ofO{(\gO)}

\global\long\def\oneo{{1,\gO}}
\global\long\def\onebdo{{1,\bndo}}
\global\long\def\info{{\infty,\gO}}

\global\long\def\infclo{{\infty,\cloo}}

\global\long\def\infbdo{{\infty,\bndo}}

\global\long\def\ld{LD}

\global\long\def\ldo{\ld\ofO}
\global\long\def\ldoo{\ldo_{0}}

\global\long\def\trace{\gamma}

\global\long\def\pr{\proj_{\rigs}}

\global\long\def\pq{\proj}

\global\long\def\qr{\,/\,\reals}

\global\long\def\aro{S_{1}}
\global\long\def\art{S_{2}}

\global\long\def\mo{m_{1}}
\global\long\def\mt{m_{2}}

\global\long\def\yieldc{B}

\global\long\def\yieldf{Y}

\global\long\def\trpr{\pi_{P}}

\global\long\def\devpr{\pi_{\devsp}}

\global\long\def\prsp{P}

\global\long\def\devsp{D}

\global\long\def\ynorm#1{\|#1\|_{\yieldf}}

\global\long\def\colls{\Psi}

\global\long\def\semib{\mathrm{SB}}

\global\long\def\tm#1{\overrightarrow{#1}}
\global\long\def\tmm#1{\underrightarrow{\overrightarrow{#1}}}

\global\long\def\itm#1{\overleftarrow{#1}}
\global\long\def\itmm#1{\underleftarrow{\overleftarrow{#1}}}

\global\long\def\ptrac{\mathcal{P}}

\global\long\def\pbase#1{\bdry_{\symm{\mi{#1}}}}

\title[Traction Hyper-Stresses]{On Jets, Almost Symmetric Tensors, and Traction Hyper-Stresses}

\author{Reuven Segev and J\k{e}drzej \'{S}niatycki }

\address{Reuven Segev\\
Department of Mechanical Engineering\\
Ben-Gurion University of the Negev\\
Beer-Sheva, Israel\\
rsegev@bgu.ac.il.}

\address{J\k{e}drzej \'Sniatycki\\
Departments of Mathematics and Statistics, University of Calgary, Calgary, Alberta, and  University of Victoria, B.C., Canada,   sniatycki@gmail.com\medskip \medskip \\}

\keywords{High-order continuum mechanics; differential geometry, vector bundle;
jet; hyper-stress; traction; multilinear algebra.}

\thanks{\today}

\subjclass[2000]{74A10; 53Z05.}
\begin{abstract}
The paper considers the formulation of higher-order continuum mechanics
on differentiable manifolds devoid of any metric or parallelism structure.
For generalized velocities modeled as sections of some vector bundle,
a variational $k$th order hyper-stress is an object that acts on jets
of generalized velocities to produce power densities. The traction
hyper-stress is introduced as an object that induces hyper-traction
fields on the boundaries of subbodies. Additional aspects of multilinear
algebra relevant to the analysis of these objects are reviewed.

\end{abstract}

\maketitle

\section{Introduction}

The present paper considers the basic mathematical objects in the
analysis of hyper-stresses for a theory defined on differentiable
manifolds. Thus, generalized velocities are represented by sections
of a vector bundle. Such a setting encompasses both the Lagrangian
and Eulerian points of views of continuum mechanics as well as classical
field theories of physics. The base manifold of the vector bundle
is interpreted accordingly as either the body manifold, the physical
space, or space-time, respectively.

As a generalization of the standard introduction of hyper-stresses
in higher-order continuum mechanics, the $k$th order hyper-stress
object, the variational hyper-stress, is dual to $k$-jets of sections
of the vector bundle (see \cite{segev_geometric_2017}). Continuum
mechanics on manifolds differs from standard formulations in Euclidean
spaces in the following significant sense. In traditional continuum
mechanics, the stress tensor plays two roles: it acts on the derivatives
of velocity fields to produce power densities and it induces traction
fields on boundaries of subbodies. For a theory on manifolds, however,
two distinct mathematical objects plays these two roles (see \cite{segev_metric-independent_2002,segev_notes_2013}).
The variational stress acts on the jets of generalized velocity fields
to produce power, while the traction stress induces the traction fields
on the boundaries of subbodies. While the variational hyper-stress
fields have been considered in \cite{segev_forces_1986,segev_geometric_2017},
we propose here a suitable candidate for the role of traction hyper-stress. 

The paper is meant to be used as an introduction to the subject and
additional details regarding symmetric tensors, used extensively in
the analysis of jets, are provided. Thus, Section \ref{sec:Jets}
introduces the basic structure, motivates the use of jets of vector
fields and describes their very basic properties. Section \ref{sec:Symmetric-Tensors-and-Jets}
considers properties of symmetric tensors of higher-order and their
use in jets, leading to the introduction of variational hyper-stresses.
Finally, Section \ref{sec:Traction-Hyper-Stresses} introduces traction
hyper-stresses and describes the basic properties of what we refer
to as ``almost symmetric tensors'' used to represent them locally.

\section{Jets\label{sec:Jets}}

Jet bundles serve as the fundamental objects in the formulation of
higher order continuum mechanics on differentiable manifolds. In this
section we review the basic constructions associated with jet bundles
of a vector bundle. Firstly, however, we motivate the use of jet bundles
in higher order continuum mechanics and classical field theories.

\subsection{The fundamental structure}

The basic object we consider here is a vector bundle 
\begin{equation}
\pi:\vb\tto\base.
\end{equation}
The object $\base$ is assumed to be a smooth manifold of dimension
$n$, that might have a boundary. We will refer to $\base$ as the
base manifold. In the context of the Lagrangian point of view of continuum
mechanics, $\base$ is interpreted as the body manifold. In the Eulerian
point of view of continuum mechanics, $\base$ is interpreted as the
physical space manifold, and in modern formulations of classical field
theories, $\base$ is interpreted as the space-time manifold.

No additional structure, such as a Riemannian metric, a connection,
a parallelism structure, is assumed for the base manifold. This level
of generality is in accordance with the reluctance of modern presentations
to use a preferred class of reference states (\eg, \cite{noll_foundations_1959}).
In particular, if one wishes to consider live tissues in bio-mechanical
studies, it is unlikely that a preferred reference state of the tissue
may be pointed out. Thus, there is no class of preferred coordinate
systems on $\base$ and denoting coordinates by $x^{i}$, $i=\oneto n$,
a coordinate transformation will be denoted by $x^{i'}=x^{i'}(x^{i})$.

Tangent vectors to the manifold $\base$ are viewed as derivatives
of curves $c:\reals\to\base$. The tangent space to $\base$ at $x$,
denoted by $T_{x}\base$ contains all the tangent vectors at $x$
and the tangent bundle $T\base$ is the collection of all tangent
vectors at the various points. Given a coordinate system $(x^{i})$
and a point $x_{0}$ with coordinates $x_{0}^{j}$, one has coordinate
lines, curves of the form $c_{i}:\reals\to\base$, such that their
coordinate representation satisfy
\begin{equation}
x^{j}(t)=x^{j}(c_{i}(t))=\begin{cases}
x_{0}^{j}, & \text{if}\;i\ne j,\\
x_{0}^{j}+t, & \text{if}\;i=j.
\end{cases}
\end{equation}
The time derivatives of these curves induce tangent vectors denoted
by $\bdry_{i}=\dot{c}_{i}$. At each point $x$, the vectors $\{\bdry_{i}\}$,
$i=\oneto n$, form a basis of $T_{x}\base$. The corresponding dual
basis of the dual vector space, $T_{x}^{*}\base$, is denoted by $\{\dee x^{i}\}$.
Thus,
\begin{equation}
\dee x^{i}(\bdry_{j})=\delta_{j}^{i}.
\end{equation}

For each $x\in\base$, $\vb_{x}:=\pi^{-1}(x)$ is a vector space that
is isomorphic to some fixed $m$-dimensional vector space $\vs$,
although no natural or particular such isomorphism is assumed. In
particular, for a pair of points $x,y\in\base$, there is no natural
isomorphism of $\vb_{x}$ with $\vb_{y}$, although both are isomorphic
to $\vs$. The mapping $\pi$ maps all vectors in $\vb_{x}$ to the
point $x$. 

Depending on the terminology and context, a vector $\vf\in\vb_{x}$
is interpreted either as a virtual velocity/displacement, or as a
generalized velocity, or as variation of the field, at the point $x$.
It should be mentioned that for the Lagrangian point of view of continuum
mechanics on manifolds, the vector bundle $\vb$ depends on the particular
configuration $\conf$ of the body in space so that $\vf$ is interpreted
as a velocity of the particle $x$ at the point $\conf(x)$ in space
or as a virtual displacement from $\conf(x)$. 

A generalized velocity field is therefore a mapping $\vf:\base\to\vb$
that assigns to each point $x$ a value for its generalized velocity.
It follows that $\pi\comp\vf=\idnt_{\base}$, \ie, $\pi(\vf(x))=x$.

A vector bundle chart, or a coordinate system, will assign to each
$\vf\in\vb$, a collection of coordinates $(x^{i},\vf^{\ga})$, where
$x^{i}$ are coordinates for the point $x=\pi(\vf)$ and $\vf^{\ga}$,
$\ga=\oneto m$, are the components of $w$ relative to some basis
$\{\sbase_{\ga}\}$ of $\vb_{x}$. It is assumed that the bases $\{\sbase_{\ga}\}$
for the various points $x$ covered by the charts depend on $x$ smoothly.
At each point $x$, covered by the charts $(x^{i},\vf^{\ga})$ and
$(x^{i'},\vf^{\ga'})$, for any $\vf\in\vb_{x}$, we must have $\vf=\vf^{\ga}\sbase_{\ga}=\vf^{\ga'}\sbase_{\ga'}$
so that there is a matrix $A_{\ga}^{\ga'}$, depending on $x$, such
that $\vf^{\ga'}=A_{\ga}^{\ga'}\vf^{\ga}$.

\subsection{Why jets}

Say $\vf:\base\to\vb$ is a velocity field. The components of $\vf(x)$
relative to the chart $(x^{i},\vf^{\ga})$ are given in terms of $m$
functions $\vf^{\ga}(x^{i})$. For the chart $(x^{i'},\vf^{\ga'})$,
the components are given by the functions $\vf^{\ga'}(x^{i'})$ and
evidently
\begin{equation}
\vf^{\ga'}(x^{i'})=A_{\ga}^{\ga'}(x^{j})\vf^{\ga}(x^{j}),
\end{equation}
where we have indicated explicitly the dependence of the matrix $A_{\ga}^{\ga'}$
on the point $x$. Differentiating the last identity, using a comma
to denote partial derivatives and the summation convention, we obtain
\begin{equation}
\vf_{,i'}^{\ga'}=A_{\ga,j}^{\ga'}x_{,i'}^{j}\vf^{\ga}+A_{\ga}^{\ga'}\vf_{,j}^{\ga}x_{,i'}^{j}.
\end{equation}
This simple relation indicates a fundamental problem. The derivatives
$\vf_{,i'}^{\ga'}$ do not depend only on the derivatives $\vf_{,i}^{\ga}$;
they depend also on the values of $\vf^{\ga}$. In other words, while
a generalized velocity as a vector field is a well-defined object,
the derivative of the generalized velocity is not a well-defined mathematical
object. One cannot separate the values of the derivatives from the
values of the velocity field in a manner that will be independent
of a chart. As an example, we observe that the derivatives may vanish
in one coordinate system while they would be different from zero in
another. Nevertheless, if we combine the values of the field and the
derivatives into a single object, the transformation rules above show
that this object\textemdash the \emph{first jet }of the generalized
velocity, $j^{1}\vf$\textemdash is well defined. Thus, the first
jet of $\vf$ is represented in the form $(x^{i},\vf^{\ga},\vf_{,j}^{\ga})$,
or we may write
\begin{equation}
j^{1}\vf=\vf^{\ga}\sbase_{\ga}+\vf_{,i}^{\ga}dx^{i}\otimes\sbase_{\ga}.
\end{equation}
The collection of $1$-jets to the vector bundle $\vb$ is denoted
as $J^{1}\vb$. 

Similarly, we may consider higher order derivatives of vector fields.
In analogy with the case of first derivatives, one realizes that under
transformation of coordinates the components of the $k$-th derivatives
$\vf_{,i'_{1}\cdots i_{'k}}^{\ga'}$depend on the values of components
of all derivatives $\vf_{,i_{1}\cdots i_{l}}^{\ga}$, $0\le l\le k$,
where we identify the zeroth derivative with the value of the function.
Thus, the invariant object is the $k$-\emph{jet} of the velocity
field represented under a coordinate system in the form 
\begin{equation}
j^{k}\vf=\vf^{\ga}\sbase_{\ga}+\vf_{,i_{1}}^{\ga}dx^{i_{1}}\otimes\sbase_{\ga}+\vf_{,i_{1}i_{2}}^{\ga}dx^{i_{1}}\otimes dx^{i_{2}}\otimes\sbase_{\ga}+\cdots+\vf_{,i_{1}\cdots i_{k}}^{\ga}dx^{i}\otimes\cdots\otimes dx^{i_{k}}\otimes\sbase_{\ga},\label{eq:Rep_Jet_NonUnique}
\end{equation}
or by $(x^{i},\vf_{\vphantom{J_{1}}}^{\ga_{\vphantom{1}}},\vf_{,j_{1}}^{\ga_{1}},\vf_{,j_{1}j_{2}}^{\ga_{2}},\dots,\vf_{,j_{1}\cdots j_{k}}^{\ga_{k}})$
. The collection of $k$-jets to $\vb$ is denoted by $J^{k}\vb$.

Since higher order continuum mechanics involves higher order derivatives
of the generalized velocities, we conclude that the terminology of
jet bundles provides an appropriate setting for the formulation of
such theories.

\subsection{Constructions involving jets}

Note that each velocity field determines a jet at any given point.
Given a chart, the representation of the jet at $x$, determined by
the velocity field $\vf$, is obtained by differentiating the components
of $\vf$ relative to the local coordinates. Any two velocity fields
will determine the same $k$-jet at $x$, if their derivatives up
to order $k$ are identical.

On the jet bundle, $J^{k}\vb$ one defines the following mappings.
The source map
\begin{equation}
\pi^{k}:J^{k}\vb\tto\base,\;\text{represented by}\;(x^{i},\vf_{\vphantom{J_{1}}}^{\ga_{\vphantom{1}}},\vf_{,j_{1}}^{\ga_{1}},\vf_{,j_{1}j_{2}}^{\ga_{2}},\dots,\vf_{,j_{1}\cdots j_{k}}^{\ga_{k}})\lmt(x^{i}),
\end{equation}
assigns to each jet the point in which it is attached. The mapping
\begin{equation}
\pi_{l}^{k}:J^{k}\vb\tto J^{l}\vb,\qquad l<k,
\end{equation}
represented by
\begin{equation}
(x^{i},\vf_{\vphantom{J_{1}}}^{\ga_{\vphantom{1}}},\vf_{,j_{1}}^{\ga_{1}},\vf_{,j_{1}j_{2}}^{\ga_{2}},\dots,\vf_{,j_{1}\cdots j_{k}}^{\ga_{k}})\lmt(x^{i},\vf_{\vphantom{J_{1}}}^{\ga_{\vphantom{1}}},\vf_{,j_{1}}^{\ga_{1}},\vf_{,j_{1}j_{2}}^{\ga_{2}},\dots,\vf_{,j_{1}\cdots j_{l}}^{\ga_{k}}),
\end{equation}
 assigns to any $k$-jet a jet of a lower order by omitting the derivatives
of order higher than $l$. In particular, identifying $J^{0}\vb$
with $\vb$, we have 
\begin{equation}
\pi_{0}^{k}:J^{k}\vb\tto\vb,
\end{equation}
which retains only the value of the generalized velocity field itself.

\section{Symmetric Tensors and Jets\label{sec:Symmetric-Tensors-and-Jets}}

As the local representation of jets involves iterated partial differentiation,
symmetric tensor are of major importance. In this section we review
the basic properties of symmetric tensors and present the way they
are used in the representation of jets.

\subsection{Multi-index notation}

Multi-index notation is very effective when high-order tensors are
involved, as is the situation here. A multi-index $\asmi I$ of length
$k$ is a $k$-tuple of positive integers, \eg, $I=\mie ik$. Multi-indices
will be denoted by upper-case roman letters and the associated indices
will be denoted by the corresponding lower case letters as in the
example above. For example, we may write the components $T_{ijk}$
of a third order tensor $T$ as $T_{\asmi I}=T_{i_{1}i_{2}i_{3}}$.
The length of a multi-index $I=\mie ik$ is denoted as the absolute
value of the multi-index, \ie,  $\abs{\asmi I}=k$. We will use the
summation convention for multi-indices so the contraction of two tensors
may be written as $T^{\asmi I}S_{\asmi I}$. When a multi-index appears
more than twice in a term, it is implied that the summation convention
for that multi-index is not in effect.

Multi-indices may be concatenated naturally so that for two multi-indices
$\asmi I$ and $\asmi J$, the concatenated multi-index is $\asmi{IJ}=\mie i{\abs I}\mie j{\abs J}$
whose length is $\abs{\asmi{IJ}}=\abs{\asmi I}+\abs{\asmi J}$. Thus,
for two tensors $S_{\asmi I}$, and $T_{\asmi J}$, one may write
$(S\otimes T)_{\asmi{IJ}}=S_{\asmi I}T_{\asmi J}$. 

For two multi-indices $\asmi{I,\;\asmi J}$, with $\abs{\asmi I}=\abs{\asmi J}=l$,
one extends the definition of the Kronecker $\gd$ by
\begin{equation}
\gd_{\asmi J}^{\asmi I}:=\gd_{j_{1}}^{i_{1}}\cdots\gd_{j_{l}}^{i_{l}}.
\end{equation}

\subsection{Symmetric tensors and permutations}

Because of the commutativity of partial derivatives that we encounter
frequently here, tensors that are completely symmetric are of particular
interest A tensor $T$ is completely symmetric if for any exchange
of two indices $i_{r}$ and $i_{s},$ $T_{i_{1}\cdots i_{r}\cdots i_{s}\cdots i_{k}}=T_{i_{1}\cdots i_{s}\cdots i_{r}\cdots i_{k}}$.

Symmetry can also be defined in terms of permutation. A permutation
of the finite ordered set $(\oneto l)$, is a bijection
\begin{equation}
\perm:(1,\dots,l)\tto(1,\dots,l).
\end{equation}
The collection of all such permutations will be denoted by $\sperm_{l}$.
From elementary combinatorics it follows that there are $l!$ permutations
in $\sperm_{l}$. For a multi-index $\asmi I$, and a permutation
$\perm$, we set 
\begin{equation}
\perm(\asmi I):=\asmi I\comp p=\miem i{\perm(l)}{\perm(1)}.
\end{equation}
Note that $i_{p(r)}$ identifies the index that arrived under the
permutation at the $r$th position, while $i_{\perm^{-1}(s)}$ is
the position of $i_{s}$ after the permutation $\perm$. Note also
that we make some abuse of notation by using the same symbol for the
permutation and its action on multi-indices. It immediately follows
that for two permutation $\perm_{1},\perm_{2}\in\sperm_{l}$,
\begin{equation}
\perm_{2}\comp\perm_{1}(\asmi I)=\asmi I\comp\perm_{1}\comp\perm_{2}.
\end{equation}
Thus, using the language of permutations, a tensor is symmetric if
for every permutation $\perm\in\sperm_{l}$, 
\begin{equation}
T_{\perm(\asmi I)}=T_{\asmi I}.
\end{equation}
\begin{rem}
We have defined symmetry above in terms of the components of the array
representing a tensor. Viewed as a multilinear mapping, a (covariant)
tensor $T$ is symmetric if 
\begin{equation}
T(v_{1},\dots,v_{l})=T(\lisubb v{\perm(1)},{\perm(l)})
\end{equation}
for any permutation $\perm$. In particular, for a symmetric tensor,
\begin{equation}
\begin{split}T_{i_{1}\cdots i_{1}} & =T(\lisubb{\sbase}{i_{1}},{i_{l}}),\\
 & =T(\lisubb{\sbase}{\perm(i_{1})},{\perm(i_{l})}),\\
 & =T_{p(i_{1})\cdots p(i_{l})}
\end{split}
\end{equation}
(see also \cite{greub_multilinear_1978}).
\end{rem}

We will use the notation $\tp^{l}\avs$ for the space of contravariant
$l$-tensors and $\stp^{l}\avs$ for the subspace of symmetric tensors.
We will also identify a tensor $T\in\tp^{l}\avs$ with the (possibly
symmetric) multilinear mapping $\avs^{*}\times\cdots\times\avs^{*}\to\reals$
in the space of (respectively, symmetric) multi-linear mappings $L^{l}(\avs^{*},\reals)$
(respectively, $L_{S}^{l}(\avs^{*},\reals)$). Thus, we make the identifications
\begin{equation}
\tp^{l}\avs\simeq L^{l}(\avs^{*},\reals),\qquad\stp^{l}\avs\simeq L_{S}^{l}(\avs^{*},\reals).
\end{equation}
The inclusion of the symmetric tensors will be denoted as 
\begin{equation}
\incl_{S}:\stp^{l}\avs\simeq L_{S}^{l}(\avs^{*},\reals)\tto\tp^{l}\avs\simeq L^{l}(\avs^{*},\reals).
\end{equation}
The analogous notation and terminology will be used for covariant
tensors.
\begin{rem}
\label{rem:Identities_Epsilon}The Levi-Civita symbol satisfies
\begin{equation}
\eps_{\asmi J}^{\asmi I}=\eps_{\mie jm}^{\mie il}=\begin{cases}
(-1)^{\perm}, & \text{if there is a permuation }\perm\text{ with }\asmi J=\perm(\asmi I),\\
0, & \text{otherwise.}
\end{cases}
\end{equation}
Thus, we set
\begin{equation}
\abs{\eps}_{\asmi J}^{\asmi I}:=\abs{\eps_{\mie jm}^{\mie il}}=\begin{cases}
1, & \text{if there is a permuation }\perm\text{ with }\asmi J=\perm(\asmi I),\\
0, & \text{otherwise.}
\end{cases}
\end{equation}
In particular,
\begin{equation}
\abs{\eps}_{\asmi J}^{\perm(\asmi I)}=\abs{\eps}_{\asmi J}^{\asmi I}.\label{eq:Epsilon_Cancels_Permutation}
\end{equation}
\end{rem}

\subsection{Cardinality sequence of a multi-index}

A multi-index $\asmi I$ induces another sequence $(I_{1},\dots,I_{n})$
in which $I_{r}$ indicates the number of times $r$ is included in
the multi-index. Evidently, $\abs{\asmi I}=\sum_{r=1}^{n}I_{r}$ and
the sequence $(I_{1},\dots,I_{n})$ is invariant under permutations
of the multi-index. 

A collection $(I_{1},\dots,I_{n})$ induces a unique non-decreasing
multi-index, \ie,  the multi-index
\begin{equation}
1\cdots12\cdots2\cdots\cdots n\cdots n
\end{equation}
where the number $r$ appears $\asmi I_{r}$ times. 

For a multi-index $\asmi I$ it is useful to write
\begin{equation}
\asmi I!=I_{1}!\cdots I_{n}!.
\end{equation}
It is observed that for a concatenated index $\asmi{IJ}$, one has
$(\asmi{IJ})_{r}=I_{r}+J_{r}$, $r=\oneto n$. The index $i=\oneto n$,
is a simple multi-index $I=i$. Obviously $\abs I=1$ and 
\begin{equation}
I_{r}=\begin{cases}
0, & \text{for }r\ne i,\\
1, & \text{for }r=i.
\end{cases}
\end{equation}
Thus, for the concatenated multi-index $\asmi Ji$, one has $(\asmi Ji)_{r}=J_{r}+\gd_{ri}$,
where $\gd$ is the Kronecker symbol.

For tensors that are symmetric with respect to a multi-index, a particular
component is indicated uniquely by a sequence in the form $(I_{1},\dots,I_{n})$
and by restricting the sequences $(i_{1},\dots,i_{\abs{\mii}})$ to
be non-decreasing. Consequently, we will use multi-indices indicated
by bold characters to be non-decreasing only and we will also write
$\mii=(I_{1},\dots,I_{n})$. In addition, the fact that a multi-index
is non-decreasing will be indicated by angle brackets, \eg,  $T_{\ordr{\asmi J}}$,
or $T_{\ordr{\mi I\mi J}}$, independently of the symmetry property
of a tensor.

\subsection{Derivatives}

Non-decreasing multi-indices are primarily used for notation involving
partial derivatives. We will use the notation
\begin{equation}
(\cdot)_{,\mii}=\bdry_{\mii}(\cdot)=\frac{\bdry^{\abs{\mii}}(\cdot)}{\partial x^{\mii}}=\frac{\bdry^{\abs{\mii}}(\cdot)}{\partial x^{i_{1}}\cdots\bdry x^{i_{\abs{\mii}}}}=\frac{\bdry^{\abs{\mii}}(\cdot)}{(\partial x^{1})^{I_{1}}\cdots(\bdry x^{n})^{I_{n}}}.
\end{equation}
Non-decreasing multi-indices may be added naturally by setting
\begin{equation}
\mii+\mi J=(I_{1}+J_{1},\dots,I_{n}+J_{n}),
\end{equation}
which determines a unique non-decreasing multi-index such that $\abs{\mii+\mi J}=\abs{\mii}+\abs{\mi J}$.
In particular,
\begin{equation}
\left((\cdot)_{,\mi I}\right)_{,\mi J}=(\cdot)_{,\ordr{\mi{IJ}}}=(\cdot)_{,\ordr{\mii+\mi J}}.
\end{equation}
Non-decreasing multi-indices can also be partially ordered so that
\begin{equation}
\mi J\le\mii,\qquad\text{if }\quad J_{r}\le I_{r},r=\oneto n.
\end{equation}
In case $\mi J\le\mii$, one can use the subtraction $\mii-\mi J$.

As hinted in the notation for partial derivatives, for $x\in\reals^{n}$,
one defines for a non-decreasing multi-index $\mii$,
\begin{equation}
x^{\mii}=(x^{1})^{I_{1}}\cdots(x^{n})^{I_{n}}.
\end{equation}

The summation convention will be applied for bold faced multi-indices,
accordingly, only to the non-decreasing sequences. For example, a
polynomial $\reals^{n}\to\reals$ of order $l$ may be written as
\begin{equation}
u=a_{\mii}x^{\mii},\qquad0\le\abs{\mii}\le l.
\end{equation}
Suspending the summation convention, it derivatives are 
\begin{equation}
u_{,\mi J}=\sum_{0\le\abs{\mii}\le l}\frac{\mii!}{(\mii-\mi J)!}x^{\mii-\mi J}.
\end{equation}
Although this relation is used mainly for the case where $J_{r}\le I_{r}$,
for all $r=\oneto n$, it may be extended to all other cases by adopting
the convention that 
\begin{equation}
\frac{1}{i!}=0,\qquad\text{for }i<0.
\end{equation}

The notation introduced above allows one to write the $l$th order
Taylor expansion of a function $f:\reals^{n}\to\reals$ in the form
\begin{equation}
\sum_{0\le\mii\le l}\frac{1}{\mii!}f_{,\mii}(x)h^{\mii}.
\end{equation}
To use the summation convention, one first sets $g_{\mii}:=f_{,\mii}/\mii!$
(no sum), and so the polynomial is written as
\begin{equation}
g_{\mii}h^{\mii},\qquad0\le\abs{\mii}\le l.
\end{equation}

\subsection{More on permutations\label{subsec:More-on-permutations}}

One observes that, for some given $\asmi I$, $\abs{\asmi I}=l$,
the sum
\begin{equation}
\sum_{\perm\in\sperm_{l}}T_{\perm(\asmi I)}
\end{equation}
contains $l!=\abs{\asmi I}!$ terms, the number of all permutations.
These include $\asmi I!$ permutations (see below) that leave $\asmi I$
invariant. In the particular case where $T$ is symmetric,
\begin{equation}
\sum_{\perm\in\sperm_{l}}T_{\perm(\asmi I)}=\abs{\asmi I}!T_{\asmi I}\inosum.
\end{equation}

On the other hand, in the expression
\begin{equation}
\abs{\eps}_{\asmi I}^{\asmi J}T_{\asmi J}=\sum_{\asmi J,\,\asmi{J=\perm(\asmi I)}}T_{\asmi J},
\end{equation}
the sum applies only to possible values of the multi-index $\asmi J$,
irrespective of the number of permutations of $\asmi I$ that give
it. Assume that $\asmi J$ is a permutation of $\asmi I$ so that
$\abs{\eps}_{\asmi I}^{\asmi J}=1$. As both $\asmi I$ and $\asmi J$
contain $I_{r}$ occurrences of the index $r$, permutations of which
leave a multi-index invariant, there are $\asmi J!=J_{1}!\cdots J_{l}!=\asmi I!$
such permutations for each $\asmi J$. Since there are $\asmi I!$
permutations that give any one particular multi-index $J$ if $\abs{\eps}_{\asmi I}^{\asmi J}\ne0$,
it follows that for any fixed $\asmi J$, 
\begin{equation}
\sum_{\substack{\perm(\asmi I)=\asmi J,\\
\perm\in\sperm_{l}
}
}T_{\perm(\asmi I)}=\asmi I!T_{\asmi J}=I!\abs{\eps}_{\asmi I}^{\asmi J}T_{\asmi J}\nosumg{\asmi{I,\,J},}\label{eq:sum_j_eq_p(i)}
\end{equation}
and so
\begin{equation}
\begin{split}\sum_{\perm\in\sperm_{l}}T_{\perm(\asmi I)} & =\sum_{\asmi J}\biggp{\sum_{\substack{\perm(\asmi I)=\asmi J,\\
\perm\in\sperm_{l}
}
}T_{\perm(\asmi I)}},\\
 & =\asmi I!\abs{\eps}_{\asmi I}^{\asmi J}T_{\asmi J}\nosumg{\asmi I.}
\end{split}
\label{eq:Epsilon_VS_permutations}
\end{equation}

We conclude that the number of non-trivial terms in the sum $\abs{\eps}_{\asmi I}^{\asmi J}T_{\asmi J}$
is 
\begin{equation}
\sum_{\asmi J}\abs{\eps}_{\asmi I}^{\asmi J}=\frac{\abs{\asmi I}!}{\asmi I!}=\frac{l!}{\asmi I!}.
\end{equation}
In the particular case where $T$ is symmetric, $\sum_{\perm\in\sperm_{l}}T_{\perm(\asmi I)}=\abs{\asmi I}!T_{\asmi I}$,
so that (\ref{eq:Epsilon_VS_permutations}) implies immediately that
\begin{equation}
\sum_{\asmi J,\,\asmi{J=\perm(\asmi I)}}T_{\asmi J}=\abs{\eps}_{\asmi I}^{\asmi J}T_{\asmi J}=\frac{\abs{\asmi I}!}{\asmi I!}T_{\asmi I}.\label{eq:epsilon_substitution}
\end{equation}

For a given pair of multi-indices, $\asmi I,\;\asmi J$, and a variable
permutation $\perm$,
\begin{equation}
\gd_{\perm(\asmi J)}^{\asmi I}=\abs{\eps}_{\asmi J}^{\asmi I}.
\end{equation}
As a result
\begin{equation}
\begin{split}\sum_{\perm\in\sperm_{l}}\gd_{\perm(\asmi J)}^{\asmi I} & =\asmi I!\abs{\eps}_{\asmi J}^{\asmi I}\nosumg{\asmi I}.\end{split}
\label{eq:Identity_delta_epsilon}
\end{equation}

\begin{rem}
For each non-decreasing multi-index $\mii$, $\abs{\mii}=l$, there
are $\abs{\mii}!/\mii!$ distinct indices $\asmi J$. Thus, the total
number of distinct multi-indices is 
\begin{equation}
\sum_{\mii}\frac{(\lisub I+n)!}{I_{1}!\cdots I_{n}!}=n^{l}
\end{equation}
\textemdash in accordance with the multinomial formula.
\end{rem}

\subsection{Symmetrization of arrays and tensors}

Any $l$-tensor $T$, having the components $T_{\asmi I}$ induces
a unique symmetric array whose components are denoted as $T_{\symm{\asmi I}}$
by 
\begin{equation}
\begin{split}T_{\symm{\asmi I}} & =\sum_{\perm\in\sperm_{l}}\frac{1}{l!}T_{\perm(\asmi I)},\\
 & =\frac{\asmi I!}{l!}\abs{\eps}_{\asmi I}^{\asmi J}T_{\asmi J}\inosum.
\end{split}
\label{eq:def-Symmetrization}
\end{equation}
We first show that $T_{\symm{\asmi I}}$ is indeed symmetric. One
has,
\begin{equation}
\begin{split}T_{\symm{q(\asmi I)}} & =\sum_{\perm\in\sperm_{l}}\frac{1}{l!}T_{\perm(q(\asmi I))},\\
 & =\sum_{\perm\in\sperm_{l}}\frac{1}{l!}T_{\perm(\asmi I)},\\
 & =T_{\symm{\asmi I}},
\end{split}
\end{equation}
where in the second line we used the fact that in the first line we
add up the terms over all permutations anyhow. One also observes that
symmetrization is a projection in the sense that the symmetrization
of a symmetric tensor yields the tensor itself. That is, if $T_{\asmi I}$
is symmetric,
\begin{equation}
\begin{split}T_{\symm{\asmi I}} & =\sum_{\perm\in\sperm_{l}}\frac{1}{l!}T_{\perm(\asmi I)},\\
 & =\sum_{\perm\in\sperm_{l}}\frac{1}{l!}T_{\asmi I},\\
 & =T_{\asmi I}.
\end{split}
\end{equation}

The symmetrization of a multi-linear mapping $T$\textemdash a covariant
tensor\textemdash is defined as the linear mapping
\begin{equation}
\smtr:\tp^{l}\avs^{*}\tto\stp^{l}\avs^{*}
\end{equation}
such that
\begin{equation}
(\smtr(T))(v_{1},\dots,v_{l})=\frac{1}{l!}\sum_{\perm\in\sperm_{l}}T(\lisubb v{\perm(1)},{\perm(l)}).\label{eq:Symmetrization_Multilinear_Mappings}
\end{equation}
In particular, 
\begin{multline}
\smtr(\sbase^{\asmi I})(v_{1},\dots,v_{l}):=\smtr(\lisupp{\sbase}{i_{1}}{\otimes}{i_{l}})(v_{1},\dots,v_{l})\\
\begin{split} & =\frac{1}{l!}\sum_{\perm\in\sperm_{l}}(\lisupp{\sbase}{i_{1}}{\otimes}{i_{l}})(v_{\perm(1)},\dots,v_{\perm(l)}),\\
 & =\frac{1}{l!}\sum_{\perm\in\sperm_{l}}(\lisupp{\sbase}{\perm(i_{1})}{\otimes}{\perm(i_{l})})(v_{1},\dots,v_{l}),
\end{split}
\label{r1}
\end{multline}
and it follows that (\emph{cf.~}\cite[p. 219]{greub_multilinear_1978})
\begin{equation}
\smtr(\sbase^{\asmi I})=\frac{1}{l!}\sum_{\perm\in\sperm_{l}}(\lisupp{\sbase}{\perm(i_{1})}{\otimes}{\perm(i_{l})})=:\lisupp{\sbase}{i_{1}}{\odot}{i_{l}}=:\stp^{\asmi I}\sbase^{\asmi I}=:\sbase^{\symm{\asmi I}}.\label{eq:Symmetrization_Tensor_products}
\end{equation}
From this definition it follows immediately that, 
\begin{equation}
\sbase_{\symm{\perm(\asmi J)}}=\sbase_{\symm{\asmi J}}\fall\perm\in\sperm.
\end{equation}
Hence, $\sbase_{\symm{\asmi J}}$ as well as all $\sbase_{\symm{\perm(\asmi J)}}$,
are represented by the non-decreasing multi-index $\mi J=\ordr{\asmi J}=(\lisub J,n)$.

Note that for a permutation $\perm\in\sperm_{l}$ and a multi-linear
mapping $T$, one may write $\perm T$ for the multi-linear mapping
defined by
\begin{equation}
(\perm T)(v_{1},\dots,v_{l}):=T(v_{\perm(1)},\dots,v_{\perm(l)}).
\end{equation}
Thus, 
\begin{equation}
\smtr(T)=\frac{1}{l!}\sum_{\perm\in\sperm_{l}}\perm T,\qquad\sbase^{\symm{\asmi I}}=\frac{1}{l!}\sum_{\perm\in\sperm_{l}}\perm\sbase^{\asmi I}=\frac{1}{l!}\sum_{\perm\in\sperm_{l}}\sbase^{\perm(\asmi I)}.
\end{equation}

The inclusion of the subspace of symmetric tensors will be denoted
by
\begin{equation}
\incl_{S}:\stp^{l}\avs^{*}\tto\tp^{l}\avs^{*}.
\end{equation}
Since the symmetrization of a symmetric tensor gives the original
tensor, the symmetrization mapping $\smtr$ is a left inverse of the
inclusion, \ie, $\smtr\comp\incl_{S}=\idnt.$ 

It is readily verified that the array $\smtr(T)_{\asmi I}$ of a symmetrized
multi-linear mapping is the symmetrized array $T_{\symm{\asmi I}}$.

\subsection{Bases and dimension}

We consider a vector space $\avs$ with some basis $\{\sbase_{i}\}$,
$i=\oneto n$. Let $T$ be a (say contravariant) tensor $T$ of degree
$l$ represented in the form
\begin{equation}
T=T^{\lisubbc i1{}l}\lisubbc{\sbase}{i_{1}}{\otimes}{i_{l}}.\label{eq:rep_Symm_Tens_NonUniq}
\end{equation}
Using multi-index notation, 
\begin{equation}
T=T^{\asmi I}\sbase_{I},\qquad\abs{\asmi I}=l,
\end{equation}
where,
\begin{equation}
\sbase_{I}:=\lisubbc{\sbase}{i_{1}}{\otimes}{i_{l}}.
\end{equation}
In particular, the dimension of the space is $n^{l}$.

The array of a symmetric tensor is uniquely determined by its components
$T^{\mii}$ for non-decreasing multi-indices only. Thus, the dimension
of the space of symmetric $l$-tensors is obviously smaller. Since
a non-decreasing $\mii$ is uniquely determined by $I_{1},\dots,I_{n}$,
the dimension may be determined accordingly. 

It is easy to realize that the number of independent component in
a symmetric $l$-tensor is $C(n+l-1,l)=(n+l-1)!/(n-1)!l!$. One considers
a string of $l$ non-decreasing indices, $I_{1}$ occurrences of $1$,
$I_{2}$ occurrences of $2$, etc., the end of each such group (except
for the last one) is indicated by a divider. Thus, the number of distinct
non-decreasing multi-indices is the number of different ways one can
place the $n-1$ (identical) dividers in the string containing $l+n-1$
elements (both indices and dividers). It follows that the dimension
of the space of symmetric $l$-tensors is $C(n+l-1,l)$.

Since a symmetric tensor is represented by a symmetric array, 
\begin{equation}
\begin{split}T & =T^{\asmi I}\sbase_{\asmi I},\\
 & =T^{\symm{\asmi I}}\sbase_{\asmi I},\\
 & =\frac{1}{l!}\sum_{\perm\in\sperm_{l}}T^{\perm(\asmi I)}\sbase_{\asmi I},\\
 & =\frac{1}{l!}\sum_{\perm\in\sperm_{l}}T^{\asmi J}\sbase_{\perm^{-1}(\asmi J)},\\
 & =\frac{1}{l!}T^{\asmi J}\sum_{q\in\sperm_{l}}\sbase_{q(\asmi J)},\\
 & =T^{\asmi J}\sbase_{(\asmi J)},
\end{split}
\end{equation}
where in the fourth line we used the fact that the order of the sum
of the multi-index and the sum over the group of permutations may
be reversed. Here, in accordance with (\ref{eq:Symmetrization_Tensor_products}),
\begin{equation}
\sbase_{(\asmi J)}=\stp^{J}\sbase_{\asmi J}:=\frac{1}{l!}\sum_{q\in\sperm_{l}}\sbase_{q(\asmi J)}.
\end{equation}
or explicitly
\begin{equation}
\sbase_{(\asmi J)}=\lisubb{\sbase}{j_{1}}{\odot}{j_{l}}:=\frac{1}{l!}\sum_{q\in\sperm_{l}}\lisubb{\sbase}{j_{q(1)}}{\otimes}{j_{q(l)}},
\end{equation}
denotes the symmetric tensor product (\emph{cf.~}\cite[p. 219]{greub_multilinear_1978}).

Furthermore,
\begin{equation}
\begin{split}T & =T^{\asmi J}\sbase_{(\asmi J)},\\
 & =\sum_{\mii}\sum_{\substack{\asmi J=\perm(\mii)\\
\perm\in\sperm_{l}
}
}T^{\asmi J}\sbase_{\asmi{\symm J}}\qquad\text{(no sum),}\\
 & =\sum_{\mii}\sum_{\substack{\asmi J=\perm(\mii)\\
\perm\in\sperm_{l}
}
}T^{\asmi J}\sbase_{(\mii)}\qquad\text{(as \ensuremath{\sbase_{(\perm(\mii))}=\sbase_{(\mii)}}),}\\
 & =\sum_{\mii}\Bigp{\sum_{\substack{\asmi J=\perm(\mii)\\
\perm\in\sperm_{l}
}
}T^{\asmi J}}\sbase_{(\mii)},\\
 & =\sum_{\mii}\frac{\abs{\mii}!}{\mii!}T^{\mii}\sbase_{\symm{\mii}}\qquad\text{(by (\ref{eq:epsilon_substitution})).}
\end{split}
\end{equation}
The last expression suggests that we make the definitions
\begin{alignat}{2}
\itm{\sbase}_{\symm{\mii}} & :=\frac{\abs{\mii}!}{\mii!}\sbase_{\symm{\mii}}, & \qquad & \itm T^{\mii}:=\frac{\abs{\mii}!}{\mii!}T^{\mii},\\
\tm{\sbase}_{\symm{\asmi J}} & :=\frac{\asmi J!}{\abs{\asmi J}!}\sbase_{\symm{\asmi J}}, &  & \tm T^{\asmi J}:=\frac{\asmi{J!}}{\abs{\asmi J}!}T^{\asmi J},
\end{alignat}
and it is noted that the fractions $\asmi J!/\abs{\asmi J}!$ are
identical for all $\asmi J=\perm(\mii)$, $\perm\in\sperm$. Utilizing
the summation convention again, we may write
\begin{gather}
T=T^{\asmi J}\sbase_{\asmi{\symm J}}=T^{\mii}\itm{\sbase}_{\symm{\mii}}=\itm T^{\mii}\sbase_{\symm{\mii}},\label{eq:rep-Symm-Tensors}\\
T^{\asmi J}\tm{\sbase}_{\symm{\asmi J}}=\tm T^{\asmi J}\sbase_{\symm{\asmi J}}=T^{\mii}\sbase_{\symm{\mii}}.
\end{gather}

Evidently, both $\{\sbase_{\symm{\mii}}\}$, and $\{\itm{\sbase}_{\symm{\mii}}\}$
are collections of linearly independent tensors and may serve as bases
for the space of symmetric tensors (\emph{cf.~}\cite[p. 219]{greub_multilinear_1978},
\cite{comon_symmetric_2008}). The components of the tensor relative
to these bases change accordingly. The representation of a symmetric
tensor in (\ref{eq:rep_Symm_Tens_NonUniq}) is in terms of regular
tensor products and is inadequate because these tensor products are
not elements of the space of symmetric tensors, in general, and because
it uses more elements than the dimension of the space. The appropriate
representation of symmetric tensors in terms of base elements is given
by (\ref{eq:rep-Symm-Tensors}).
\begin{example}
\label{exa:Matrix_of_inclusion}We consider now the inclusion
\begin{equation}
\incl_{S}:\stp^{l}\avs\tto\tp^{l}\avs.
\end{equation}
The matrix of the inclusion relative to the bases $\sbase_{\asmi J}$
in $\tp^{l}\avs$ and $\itm{\sbase}_{\symm{\mii}}$ in $\stp^{l}\avs$
satisfies
\begin{equation}
\begin{split}(\incl_{S})_{\mii}^{\asmi J}\sbase_{\asmi J} & =\incl_{S}(\itm{\sbase}_{\symm{\mii}}),\\
 & =\frac{\abs{\mii}!}{\mii!}\incl_{S}(\sbase_{\symm{\mii}}),\\
 & =\frac{\abs{\mii}!}{\mii!}\sbase_{\symm{\mii}},\\
 & =\frac{1}{\mii!}\sum_{\perm\in\sperm_{l}}\sbase_{\perm\symm{\mii}},\qquad\text{no sum on }\mii,\\
 & =\frac{1}{\mii!}\sum_{\perm\in\sperm_{l}}\gd_{\perm(\mii)}^{\asmi J}\sbase_{\asmi J},\qquad\text{no sum on }\mii,\\
 & =\frac{\mii!}{\mii!}\abs{\eps}_{\mii}^{\asmi J}\sbase_{\asmi J},\qquad\text{using (\ref{eq:Identity_delta_epsilon}).}
\end{split}
\end{equation}
It is concluded that
\begin{equation}
(\incl_{S})_{\mii}^{\asmi J}=\abs{\eps}_{\mii}^{\asmi J}.
\end{equation}
In addition, as the components of $T$ relative to the basis $\{\itm{\sbase}_{\symm{\mii}}\}$
are $T^{\mii}$, 
\begin{equation}
\begin{split}\incl_{S}(T) & =(\incl_{S})_{\mii}^{J}T^{\mii}\sbase_{\asmi J},\\
 & =\abs{\eps}_{\mii}^{\asmi J}T^{\mii}\sbase_{\asmi J},
\end{split}
\end{equation}
or,
\begin{equation}
T^{\asmi J}=(\incl_{S}(T))^{\asmi J}=\abs{\eps}_{\mii}^{\asmi J}T^{\mii},
\end{equation}
which could have been deduced otherwise.
\end{example}

\begin{example}
\label{exa:Matrix_of_Symmetrization}Consider the symmetrization mapping
$\smtr:\tp^{l}\avs\to\stp^{l}\avs$. One has,
\begin{equation}
\begin{split}\smtr(\sbase_{\asmi J}) & :=\sbase_{(\asmi J)},\\
 & \hphantom{:}=\abs{\eps}_{\asmi J}^{\mii}\sbase_{\symm{\mii}}\commen{(only\,one\,}\mii\text{),}\\
 & \hphantom{:}=\abs{\eps}_{\asmi J}^{\mii}\frac{\asmi{\mii!}}{\abs{\mii}!}\itm{\sbase}_{\symm{\mii}},
\end{split}
\end{equation}
and it follows from the definition of a matrix that
\begin{equation}
\smtr_{\asmi J}^{\mii}=\frac{\asmi{\mii!}}{\abs{\mii}!}\abs{\eps}_{\asmi J}^{\mii}.
\end{equation}
In addition,
\begin{equation}
\begin{split}\smtr(T)^{\mii} & =\smtr_{\asmi J}^{\mii}T^{\asmi J},\\
 & =\frac{\asmi{\mii!}}{\abs{\mii}!}\abs{\eps}_{\asmi J}^{\mii}T^{\asmi J},\\
 & =T^{(\mii)},\commen{usings\,(\ref{eq:def-Symmetrization}).}
\end{split}
\end{equation}
\end{example}

\subsection{Duality}

Consider the dual basis $\{\sbase^{i}\}$ of the dual vector space
$\avs^{*}$ so that $\sbase^{i}(\sbase_{j})=\gd_{j}^{i}$. For any
two multi-indices $\asmi I,\asmi J$, with $\abs{\asmi I}=\abs{\asmi J}=l$,
we consider the action $\sbase^{\symm{\asmi I}}(\sbase_{\symm{\asmi J}})$.
We have
\begin{equation}
\begin{split}\sbase^{\symm{\asmi I}}(\sbase_{\symm{\asmi J}}) & =(\lisuppc{\sbase}{i_{1}}{\odot}{i_{l}})(\lisubbc{\sbase}{j_{1}}{\odot}{j_{l}}),\\
 & =\frac{1}{(l!)^{2}}\biggp{\sum_{\perm\in\sperm_{l}}\lisuppc{\sbase}{i_{p(1)}}{\otimes}{i_{p(l)}}}\biggp{\sum_{q\in\sperm_{l}}\lisubbc{\sbase}{j_{q(1)}}{\otimes}{j_{q(l)}}},\\
 & =\frac{1}{(l!)^{2}}\sum_{\perm\in\sperm_{l}}\biggp{\sum_{q\in\sperm_{l}}\gd_{q(\asmi J)}^{\perm(\asmi I)}},\\
 & =\frac{1}{(l!)^{2}}\sum_{\perm\in\sperm_{l}}\asmi I!\abs{\eps}_{\asmi J}^{\perm(\asmi I)}\qquad\text{(using Equation (\ref{eq:Identity_delta_epsilon})),}\\
 & =\frac{\asmi I!}{(l!)^{2}}\sum_{\perm\in\sperm_{l}}\abs{\eps}_{\asmi J}^{\asmi I}\qquad\hphantom{xxx}\text{(using Equation (\ref{eq:Epsilon_Cancels_Permutation})),}\\
 & =\frac{\asmi I!l!}{(l!)^{2}}\abs{\eps}_{\asmi J}^{\asmi I}\qquad\hphantom{xxxxxix}\text{(there are }l!\text{ permutations),}\\
 & =\frac{\asmi{I!}}{\abs{\asmi I}!}\abs{\eps}_{\asmi J}^{\asmi I}.
\end{split}
\end{equation}
It follows from the identity above that for non-decreasing multi-indices
$\mii,\mi J$,
\begin{equation}
\begin{split}\sbase^{\symm{\mi I}}(\itm{\sbase}_{\symm{\mi J}}) & =\sbase^{\symm{\mi I}}\biggp{\frac{\abs{\mi J}!}{\mi J!}\sbase_{\symm{\mi J}}},\\
 & =\abs{\eps}_{\mi J}^{\mi I},\\
 & =\gd_{\mi J}^{\mii},
\end{split}
\label{eq:Dual_Basis_Symm_Tens}
\end{equation}
where one realizes that if the two multi-indices are non-decreasing,
one can be a permutation of the other only when they are equal.

The last identity implies that the basis $\{\sbase^{\symm{\mii}}\}$
is the dual basis of $\{\itm{\sbase}_{\symm{\mi J}}\}$, and in particular,
\begin{equation}
\stp^{l}\avs^{*}\simeq\left(\stp^{l}\avs\right)^{*}.
\end{equation}
Finally, for $T=T^{\mii}\itm{\sbase}_{\symm{\mii}}\in\stp^{l}\avs$,
and $\psi=\psi_{\mii}\sbase^{\symm{\mii}}\in\stp^{l}\avs^{*}$,
\begin{equation}
\psi(T)=\psi_{\mii}T^{\mii}.\label{eq:Action_of_Cotensors}
\end{equation}

\subsection{Symmetrization of co-tensors and co-symmetrization}

The inclusion of symmetric tensors in the collection of all tensors
induces by duality a projection 
\begin{equation}
\incl_{S}^{*}:\left(\tp^{l}\avs\right)^{*}\simeq\tp^{l}\avs^{*}\tto\left(\stp^{l}\avs\right)^{*}\simeq\stp^{l}\avs^{*},
\end{equation}
such that 
\begin{equation}
\incl_{S}^{*}(\vph)(T)=\vph(\incl(T)),
\end{equation}
for every symmetric tensor $T$. Thus, referring to elements of $\left(\tp^{l}\avs\right)^{*}$
as \emph{co-tensors}, $\incl_{S}^{*}$ is a symmetrization operator
for co-tensors.

One obtains
\begin{equation}
\begin{split}(\incl_{S}^{*}(\vph))_{\mii} & =(\incl_{S}^{*})_{\mii}^{\asmi J}\vph_{\asmi J},\\
 & =\abs{\eps}_{\mii}^{\asmi J}\vph_{\asmi J},
\end{split}
\label{eq:projection_of_cotensors}
\end{equation}
where we observe that in the last expression one adds up the components
of $\vph$ corresponding to all permutations of $\mii$, similarly
to the symmetrization operation (but without taking the average). 

In addition,
\begin{equation}
\begin{split}\incl_{S}^{*}(\vph)(T) & =(\incl_{S}^{*}(\vph))_{\mii}T^{\mii},\\
 & =\abs{\eps}_{\mii}^{\asmi J}\vph_{\asmi J}T^{\mii},\\
 & =\vph_{\asmi J}T^{\asmi J},
\end{split}
\end{equation}
as expected. In the particular case where $\vph$ is symmetric, using
(\ref{eq:epsilon_substitution}), (\ref{eq:projection_of_cotensors})
gives
\begin{equation}
(\incl_{S}^{*}(\vph))_{\mii}=\abs{\eps}_{\mii}^{\asmi J}\vph_{\asmi J}=\frac{\abs{\mi I}!}{\mi I!}\vph_{\mii},
\end{equation}
and
\begin{equation}
\incl_{S}^{*}(\vph)(T)=\vph_{\asmi J}T^{\asmi J}=\sum_{\mii}\frac{\abs{\mi I}!}{\mi I!}\vph_{\mii}T^{\mii}.
\end{equation}

The dual of the symmetrization mapping is (the co-symmetrization)
\begin{equation}
\smtr^{*}:\stp^{l}\avs^{*}\tto\tp^{l}\avs^{*},
\end{equation}
given by 
\begin{equation}
\smtr^{*}(\psi)(T)=\psi(\smtr(T)).
\end{equation}
Using the matrix obtained in Example \ref{exa:Matrix_of_Symmetrization},
\begin{equation}
\begin{split}\smtr^{*}(\psi)(T) & =(\smtr^{*})_{\asmi J}^{\mii}\psi_{\mii}T^{\asmi J},\\
 & =\sum_{\mii,\asmi J}\frac{\mii!}{\abs{\mii}!}\abs{\eps}_{\asmi J}^{\mii}\psi_{\mii}T^{\asmi J},
\end{split}
\end{equation}
and it follows that
\begin{equation}
\smtr^{*}(\psi)_{\asmi J}=\sum_{\mii}\frac{\mii!}{\abs{\mii}!}\abs{\eps}_{\asmi J}^{\mii}\psi_{\mii}.
\end{equation}
(It is observed that the sum over $\mii$ contains only one non-trivial
term.) In other words, if $\asmi J$ is a permutation of $\mii$,
then, $\mii=\ordr J$ ($\mii$ is obtained by ordering $\asmi J$),
and 
\begin{equation}
\smtr^{*}(\psi)_{\asmi J}=\frac{\mii!}{\abs{\mii}!}\psi_{\mii}=\frac{\asmi J!}{\abs{\asmi J}!}\psi_{\ordr{\asmi J}}.
\end{equation}
In particular, it $T$ is symmetric, $\smtr^{*}(\psi)(T)=\psi(\smtr(T))=\psi(T)$,
and so
\begin{equation}
\sum_{\asmi J}\frac{\asmi J!}{\abs{\asmi J}!}\psi_{\ordr{\asmi J}}T^{\asmi J}=\sum_{\mii}\psi_{\mii}T^{\mii}.
\end{equation}
The last equation simply implies that for each non-decreasing $\mii$
there are $\abs{\mii}!/\mii!=\abs{\asmi J}!/\asmi J!$ distinct indices
$\asmi J$ obtained by permutations.

Setting
\begin{equation}
\itmm T^{\mii}:=\frac{\abs{\mii}!}{\mii!}T^{\mii},\qquad\tmm T^{K}:=\frac{K!}{\abs K!}T^{K},\label{eq:T-mmod-defn}
\end{equation}
one can write
\begin{equation}
\psi_{\ordr{\asmi J}}\tmm T^{\asmi J}=\sum_{\mii}\psi_{\mii}T^{\mii},\qquad\psi_{\ordr{\asmi J}}T^{\asmi J}=\psi_{\mii}\itmm T^{\mii}.\label{eq:T-mmod-prop}
\end{equation}

\subsection{Application to jets}

We want to use the notation introduced above to represent, locally,
elements of jet bundles. The tensors considered above are homogeneous
in the sense that they have a definite order, a local representation
of a $k$-jet is an element of the symmetric algebra and is represented
in general by a collection of symmetric tensors of all orders $l\le k$.
We recall that the representation in (\ref{eq:Rep_Jet_NonUnique})
uses the regular tensor products that are not appropriate base vectors. 

The multi-linear mappings that represent a jet are not real valued.
Rather, they are valued in $\avs$\textemdash the typical fiber of
the vector bundle. We use a local basis $\{\sbase_{\ga}\}$ for the
vector spaces $\vb_{x}$ so that a section of $\vb$ is locally of
the form
\begin{equation}
\vf=\vf^{\ga}\sbase_{\ga},
\end{equation}
where the components $\vf^{\ga}$ are real valued functions. This
does not affect the symmetry properties considered above. The basic
vector space on which the tensors are defined at each point is the
tangent space of the manifold at that point. Given a chart with coordinates
$(x^{i})$, the base vectors induced are $\{\bdry_{i}\}$ and they
replace the base vectors $\{\sbase_{i}\}$ used above. The various
derivatives in $\vf_{,\mii}^{\ga}$ are covariant tensors and are
represented using the dual basis $\{\dee x^{i}\}$. The derivatives
$\vf_{,\mii}^{\ga}(x)$, $\abs{\mii}=l$, are elements of
\begin{equation}
L_{S}^{l}(T_{x}\base,\vb_{x})\simeq\stp^{l}T_{x}^{*}\base\otimes\vb_{x}
\end{equation}

Thus, we may rewrite now (\ref{eq:Rep_Jet_NonUnique}) in the form
\begin{equation}
j^{k}\vf=\vf_{,\mii}^{\ga}\itm{\dee x}^{\symm{\mii}}\otimes\sbase_{\ga}=\itm{\vf}_{,\mii}^{\ga}\dee x^{\symm{\mii}}\otimes\sbase_{\ga},\qquad0\le\abs{\mii}\le k.
\end{equation}
An element $A\in J^{k}\vb$ of the jet bundle with $\pi^{k}(A)=x\in\base$
is of the form
\begin{equation}
A=j_{x}^{k}\vf:=(j^{k}\vf)(x),
\end{equation}
for some section $\vf$ which may be represented locally as 
\begin{equation}
A=\vf_{,\mii}^{\ga}(x)\itm{\dee x}^{\symm{\mii}}\otimes\sbase_{\ga}.\label{eq:rep_jet_element_inc_ind}
\end{equation}
Noting that the values of the various $\vf_{,\mii}^{\ga}(x)$ are
not constrained by compatibility, any element of the jet bundle may
be represented in the form
\begin{equation}
A=A_{\mii}^{\ga}\itm{\dee x}^{\symm{\mii}}\otimes\sbase_{\ga}=\itm A_{\mii}^{\ga}\dee x^{\symm{\mii}}\otimes\sbase_{\ga},\qquad0\le\abs{\mii}\le k,
\end{equation}
$A_{\mii}^{\ga}\in\stp^{\abs{\mii}}T_{x}^{*}\base\otimes\vb_{x}$.
Given an element of the jet bundle, one can construct a local section
representing it by using the corresponding Taylor polynomial in any
chart. 

We finally remark that the representation using $\itm{\dee x}^{\symm{\mii}}$seems
preferable because the components of the jet are the exactly the derivatives.

\subsection{Duality for jets\label{subsec:Duality-for-jets}}

In view of (\ref{eq:Dual_Basis_Symm_Tens}), the dual basis of $\{\itm{\dee x}^{\symm{\mii}}\mid0\le\abs{\mii}\le k\}$
is $\{\pbase I\mid0\le\abs{\mii}\le k\}$. Note that $\pbase I:=\bdry_{i_{1}}\odot\cdots\odot\bdry_{i_{\abs{\mii}}}$
is the symmetrized tensor product while $\bdry_{\mii}$ is the differential
operator which is symmetric automatically. Real valued linear mappings
on the space of jets at a point $x\in\base$ make up the dual space
$(J_{x}^{k}\vb)^{*}$. Such a linear functional 
\begin{equation}
\vph:J_{x}^{k}\vb\tto\reals
\end{equation}
is locally of the form
\begin{equation}
\vph=\vph_{\ga}^{\mii}\pbase I\otimes\sbase^{\ga},
\end{equation}
so that for $\vph\in(J_{x}^{k}\vb)^{*}$, $A=j^{k}\vf(x)\in J_{x}^{k}\vb$,
\begin{equation}
\vph(A)=\vph_{\ga}^{\mii}A_{\mii}^{\ga}=\vph_{\ga}^{\mii}\vf_{,\mii}^{\ga},
\end{equation}
where $0\le\abs{\mii}\le k$, unless indicated otherwise. 

\subsection{Variational hyper-stresses\label{subsec:Variational-hyper-stresses}}

In accordance with the variational approach to higher order continuum
mechanics, we view variational hyper-stresses as fields that act on
the derivatives of the virtual velocities to produce power densities
(see \cite{segev_geometric_2017}). Thus, in the current setting,
a variational hyper-stress object should act linearly on the $k$-jet
of a field $\vf$ to produce a density on $\base$.

We recall that for integration over an $n$-dimensional manifold,
such as $\base$, densities (integrands) are $n$-forms\textemdash alternating
(completely anti-symmetric) tensor fields of order $n$. The space
of $r$-alternating tensors over $T_{x}\base$ will be denoted by
$\ext^{r}T_{x}^{*}\base$ and the bundle of alternating tensors is
$\ext^{r}T^{*}\base$. A local coordinate system $(x^{i})$, induces
such an $n$-form
\begin{equation}
\dee x=\lisuppc{\dee x}1{\wedge}n,\label{eq:Volume_element}
\end{equation}
where a wedge denotes the exterior product\textemdash the anti-symmetrized
tensor product. Note that anti-symmetric tensors cannot have repeated
indices and so the multi-indices representing base vectors and components
are strictly increasing rather than non-decreasing. This implies that
$\ext^{n}T_{x}^{*}\base$ is one dimensional, and for which $\dee x$,
induced by a local coordinate system, may serve as a basis. Thus,
every $n$-form may be written locally as
\begin{equation}
\theta=\vartheta(x)\dee x
\end{equation}
for a real valued function $\vartheta$. 

In view of these observations, a variational hyper-stress object at
$x$ should be a linear mapping
\begin{equation}
\std_{x}:J_{x}^{k}\vb\tto\ext^{r}T_{x}^{*}\base
\end{equation}
so that $\std_{x}(j^{k}\vf(x))$ is the power density. Denoting the
bundle of linear mappings $J^{k}\vb\to\ext^{n}T^{*}\base$ by $\L{J^{k}\vb,\ext^{n}T^{*}\base}$,
\begin{equation}
\std_{x}\in\L{J_{x}^{k}\vb,\ext^{n}T_{x}^{*}\base}=\L{J^{k}\vb,\ext^{n}T^{*}\base}_{x}.
\end{equation}
It is also observed that 
\begin{equation}
\L{J_{x}^{k}\vb,\ext^{n}T_{x}^{*}\base}=(J_{x}^{k}\vb)^{*}\otimes\ext^{n}T_{x}^{*}\base,
\end{equation}
and 
\begin{equation}
\L{J^{k}\vb,\ext^{n}T^{*}\base}=(J^{k}\vb)^{*}\otimes_{\base}\ext^{n}T^{*}\base.
\end{equation}

We conclude that a variational hyper-stress field is a section $\std$
of $\L{J^{k}\vb,\ext^{n}T^{*}\base}$. In view of the representation
of elements of the dual to the jet bundle in Section \ref{subsec:Duality-for-jets},
the local representation of $\std$ is of the form
\begin{equation}
\std=\std_{\ga}^{\mii}\pbase I\otimes\sbase^{\ga}\otimes\dee x.
\end{equation}
The action of a variational hyper-stress on the jet of a generalized
velocity is the density given by
\begin{equation}
\std(j^{k}\vf)=\std_{\ga}^{\mii}\vf_{,\mii}^{\ga}\dee x
\end{equation}
and the total power is
\begin{equation}
P=\int_{\base}\std\cdot j^{k}\vf,
\end{equation}
where $\std\cdot j^{k}\vf$ is the $n$-form $(\std\cdot j^{k}\vf)(x)=\std(x)(j^{k}\vf(x))$.

\section{Traction Hyper-Stresses and Almost Symmetric Tensors\label{sec:Traction-Hyper-Stresses}}

The stress object in traditional continuum mechanics plays two roles.
On the one hand, from the variational point of view, the stress object
acts on the derivative of the velocity field to produce power. The
generalization of this object is the variational hyper-stress introduced
above. On the other hand, as a result of Cauchy's stress theorem,
the stress object determines the traction field on the boundary of
the body and its sub-bodies. While the same mathematical object plays
these two roles in the traditional formulation, in the case of a formulation
on manifolds, the traction is determined by a different mathematical
object\textemdash the \emph{traction stress} (see \cite{segev_notes_2013}).

\subsection{Traction and traction stresses}

For the case $k=1$\textemdash first order continuum mechanics\textemdash the
traction field on the boundary of $\base$, or in general, any of
its sub-bodies (sub-regions) $\reg$, acts linearly on the values
of the generalized velocity $\vf$ to produce a power density over
the boundary, the flux of power. Since the boundaries are manifolds
of dimensions $n-1$, a power density over the boundary $\bdry\reg$
is an $(n-1)$-form over $\bdry\reg$, that is, a section of $\ext^{n-1}T^{*}\bdry\reg$.
Thus, the traction field on the boundary is a section of 
\begin{equation}
\L{\vb,\ext^{n-1}T^{*}\bdry\reg},
\end{equation}
where, with some abuse of notation, we have omitted the indication
that we restrict $\vb$ to $\bdry\reg$. It is observed that the fibers
of $\ext^{n-1}T^{*}\bdry\reg$ are $1$-dimensional.

A traction stress\textemdash an object that unlike a traction field
is defined over the entire $\base$\textemdash should induce a traction
field on the boundary of each sub-region using a generalization of
Cauchy's formula. A natural candidate for such a mathematical object
is suggested by the following observation. While the space of $(n-1)$-alternating
tensors over $\bdry\reg$ is $1$-dimensional, the space $\ext^{n-1}T^{*}\base$
of $(n-1)$-alternating tensors over $\base$ is $n$-dimensional.
While an element of $\ext^{n-1}T^{*}\base$ assigns a value to any
collection of $n-1$ vectors, an element of $\ext^{n-1}T^{*}\bdry\reg$
assigns values only to vectors tangent to $\bdry\reg$. In fact, an
element of $\ext^{n-1}T^{*}\base$ may be restricted to act on vectors
tangent to $\bdry\reg$ for every sub-body $\reg$. Thus, for each
sub-body $\reg$, we have a restriction mapping
\begin{equation}
\rho_{\bdry\reg}:\ext^{n-1}T^{*}\base\tto\ext^{n-1}T^{*}\bdry\reg,
\end{equation}
naturally defined by 
\begin{equation}
\rho_{\partial\reg}(\go)(\lisubb v1,n)=\go(\lisubb v1,n),\qquad v_{r}\in T\bdry\reg.
\end{equation}

Thus, a traction stress is defined to be an element 
\begin{equation}
\tst_{0}\in\L{\vb,\ext^{n-1}T^{*}\base}.
\end{equation}
Given a traction stress $\tst_{0}$, at a point $x$, for any sub-body
$\reg$ with $x\in\bdry\reg$, a traction $\sfc_{0}\in\L{\vb,\ext^{n-1}T^{*}\bdry\reg}$
is determined at $x$ be setting
\begin{equation}
\sfc_{0}=\wh{\rho}_{\bdry\reg}(\tst)=\rho_{\bdry\reg}\comp\tst,\quad\text{\emph{i.e., }\quad}\sfc_{0}(\vf)=\rho_{\partial\reg}(\tst(\vf)).\label{eq:Cauchy_formula_stresses}
\end{equation}
The last equation is the required generalization of Cauchy's formula
to the setting of differentiable manifolds. In analogy with the classical
Cauchy theorem, it can be shown that if the traction is given on the
boundary of every sub-body $\reg$, with $x\in\bdry\reg$, then, assuming
certain consistency conditions hold, a unique traction stress is determined
at $x$ (see \cite{segev_cauchys_1999,segev_notes_2013} for details).

A traction stress field is a section of the bundle $\L{\vb,\ext^{n-1}T^{*}\base}$. 

\subsection{On the local representation of $(n-1)$-forms and traction stresses}

Traction stresses are elements of 
\begin{equation}
\L{\vb,\ext^{n-1}T^{*}\base}\simeq\vb^{*}\otimes\ext^{n-1}T^{*}\base.
\end{equation}
Thus, we make a few comments on the representation of $(n-1)$-alternating
tensors, \ie, for a vector space, $\avs$, we consider elements of
$\ext^{n-1}\avs^{*}$.

We first recall that $\ext^{n}\avs^{*}$ is one-dimensional and that
$\ext^{n-1}\avs^{*}$ is $n$-dimensional. Let $\contr$ denotes the
contraction (inner product) whereby for an alternating $r$-tensor
$\go\in\ext^{r}\avs^{*}$ and a vector $v_{1}\in\avs$, $v_{1}\contr\go$
is the alternating $(r-1)$-tensor such that
\begin{equation}
v_{1}\contr\go(\lisubb v2,r)=\go(\lisub v,r).
\end{equation}
In fact, considering the particular case $r=n-1$, one can view the
contraction as a mapping
\begin{equation}
\wh{\contr}:\avs\times\ext^{n}\avs^{*}\tto\ext^{n-1}\avs^{*},\qquad\wh{\contr}(v,\theta)=v\contr\theta.
\end{equation}
We observe that the definition of the contraction mapping implies
immediately that the mapping, $\wh{\contr}$ is bi-linear. It follows
from the universality property of tensor products that there is a
linear mapping, which we still denote as $\wh{\contr}$, such that
\begin{equation}
\wh{\contr}:\avs\otimes\ext^{n}\avs^{*}\tto\ext^{n-1}\avs^{*},\qquad\wh{\contr}(v\otimes\theta)=v\contr\theta.
\end{equation}
One can verify that this mapping is injective (\eg, \cite{segev_notes_2013}),
and as the dimensions match, it follows that $\wh{\contr}$ defines
a natural isomorphism
\begin{equation}
\avs\otimes\ext^{n}\avs^{*}\simeq\ext^{n-1}\avs^{*}.
\end{equation}
Furthermore, for a basis $\{\sbase_{i}\}$, a natural basis of $\ext^{n}\avs^{*}$
is $\lisuppc{\sbase}1{\wedge}n$, and so 
\begin{equation}
\{\sbase_{i}\contr(\lisuppc{\sbase}1{\wedge}n)\},\qquad i=\oneto n,
\end{equation}
may serve as a natural basis to $\ext^{n-1}\avs^{*}$.

Going back to traction stresses, it follows from the foregoing discussion,
that
\begin{equation}
\L{\vb,\ext^{n-1}T^{*}\base}\simeq\vb^{*}\otimes\ext^{n-1}T^{*}\base\simeq\vb^{*}\otimes T\base\otimes\ext^{n}T^{*}\base.
\end{equation}
For a given coordinate system $(x^{i})$, the collection $\{\bdry_{i}\contr\dee x\}$
may serve as a basis for $(\ext^{n-1}T^{*}\base)_{x}$. As a result,
any $\go$ may be represented locally in the form
\begin{equation}
\go=\go^{i}\bdry_{i}\contr\dee x,
\end{equation}
where $\dee x$ is defined in (\ref{eq:Volume_element}). The local
representation of a traction stress will be
\begin{equation}
\tst=\tst_{\ga}^{i}\sbase^{\ga}\otimes(\bdry_{i}\contr\dee x)
\end{equation}
and 
\begin{equation}
\tst(\vf)=\tst_{\ga}^{i}\vf^{\ga}(\bdry_{i}\contr\dee x).
\end{equation}

\subsection{Hyper-traction and traction hyper-stresses}

By analogy with the case $k=1$ described above, where the traction
object acts on the $k-1=0$-jet of the generalized velocity, we propose
that a \emph{hyper-traction} on the boundary $\bdry\reg$ of a sub-body
$\reg$, be defined as an element
\begin{equation}
\sfc\in\L{J^{k-1}\vb,\ext^{n-1}T^{*}\bdry\reg}\simeq(J^{k-1}\vb)^{*}\otimes\ext^{n-1}T^{*}\bdry\reg.
\end{equation}
Thus, the total power flux is given by 
\begin{equation}
\int_{\bdry\reg}\sfc\cdot j^{k-1}\vf.
\end{equation}

A traction hyper-stress field is defined in analogy with the definition
of a traction stress, in the sense that it acts on a lower order jet
to give an $(n-1)$-form which can be integrated on the boundaries
of sub-bodies. Thus, a \emph{traction hyper-stress} is defined to
be an element 
\begin{equation}
\st_{0}\in\L{J^{k-1}\vb,\ext^{n-1}T^{*}\base}\simeq(J^{k-1}\vb)^{*}\otimes T\base\otimes\ext^{n}T^{*}\base.
\end{equation}
It follows from the foregoing analysis that a traction hyper-stress
is represented locally in the form
\begin{equation}
\st_{0}=\tst_{\ga}^{\mi Jj}\pbase J\otimes\sbase^{\ga}\otimes(\bdry_{i}\contr\dee x),\qquad0\le\abs{\mi J}\le k-1.
\end{equation}
A traction hyper-stress field is a section of $\L{J^{k-1}\vb,\ext^{n-1}T^{*}\base}$
and the action of a hyper-stress field $\tst$ on the $(k-1)$-jet
of a generalized velocity $\vf$ is given by
\begin{equation}
\tst\cdot j^{k-1}\vf=\st_{\ga}^{\mi Jj}\vf_{,\mi J}^{\ga}\bdry_{i}\contr\dee x.
\end{equation}

These natural extensions imply that the Cauchy formula (\ref{eq:Cauchy_formula_stresses})
remains applicable as it simply represents is the restriction of forms.
Thus, given a traction hyper-stress field $\tst$, and a generalized
velocity field $\vf$, the total flux of power through the boundary
$\bdry\reg$ is 
\begin{equation}
\int_{\bdry\reg}\sfc\cdot j^{k-1}\vf=\int_{\bdry\reg}\wh{\rho}_{\bdry\reg}(\st)\cdot j^{k-1}\vf.
\end{equation}

It is emphasized that the array $\st_{\ga}^{\mi Jj}$ representing
a traction hyper-stress, is symmetric with respect to permutations
of the multi-index $\mi J$ and for this reason it appears in conjunction
with the symmetrized basis $\pbase J$. In particular, no symmetry
is expected for permutations that ``mix'' the indices $\mi J$ and
$j$. Thus, for a fixed value $l=\abs{\mi J}$, we refer to the tensor
$\tst_{\ga}^{\mi Jj}$ as \emph{almost symmetric tensor}. 

\subsection{Almost symmetric tensors}

In order to simplify the notation we will consider henceforth only
real valued almost symmetric tensors. That is, for some given vector
space, $\avs$, we consider elements of $\bigp{\stp^{l-1}\avs}\otimes\avs$
rather than elements of $\bigp{\stp^{l-1}\avs}\otimes\avs^{*}\otimes\avs\otimes\ext^{n}\avs^{*}$. 

Let $\{\sbase_{i}\}$ be a basis in $\avs$. Then, we may use either
$\{\sbase_{\symm{\mi J}}\}$, $0\le\abs{\mi J}\le l-1$ or the basis
$\{\itm{\sbase}_{\symm{\mi J}}\}$ for $\stp^{l-1}\avs$ in analogy
with (\ref{eq:rep-Symm-Tensors}). A real valued almost symmetric
tensor $T$ can be represented in the form
\begin{equation}
T=T^{\asmi I}\sbase_{\asmi I}=T^{\mi Jj}\itm{\sbase}_{\symm{\mi J}}\otimes\sbase_{j}=\itm T^{\mi Jj}\sbase_{\symm{\mi J}}\otimes\sbase_{j},
\end{equation}
where $0\le\abs{\mi J}\le l-1,\;0\le\abs{\asmi I}\le l$ and
\begin{equation}
\itm{\sbase}_{\symm{\mi J}}=\frac{(l-1)!}{\mi J!}\sbase_{\symm{\mi J}},\qquad\itm T^{\mi J}=\frac{(l-1)!}{\mi J!}T^{\mi J}.
\end{equation}

For the dual space we have
\begin{equation}
\left[\bigp{\stp^{l-1}\avs}\otimes\avs\right]^{*}\simeq\bigp{\stp^{l-1}\avs^{*}}\otimes\avs^{*}
\end{equation}
so that its elements may be referred to as almost symmetric co-tensors.
For the basis $\{\itm{\sbase}_{\symm{\mi J}}\otimes\sbase_{j}\}$,
the dual basis will be $\{\sbase^{\symm{\mi J}}\otimes\sbase^{j}\}$.
An element $\vph$ of $\left[\bigp{\stp^{l-1}\avs}\otimes\avs\right]^{*}$
is represented in the form
\begin{equation}
\vph=\vph_{\mi Jj}\sbase^{\symm{\mi J}}\otimes\sbase^{j}
\end{equation}
with
\begin{equation}
\vph(T)=\vph_{\mi Jj}T^{\mi Jj}.
\end{equation}

\section{Conclusion}

We have reviewed above the language needed for the formulation of
higher order continuum mechanics on differentiable manifolds. In particular,
we have proposed the mathematical object that we believe should play
the role of traction hyper-stress. While for the case $k=1$, the
traction stress has been defined in \cite{segev_metric-independent_2002,segev_notes_2013},
no natural analogous definition has been presented in \cite{segev_geometric_2017}.
In fact, in \cite{segev_geometric_2017} some of the difficulties
have been indicated and subsequently avoided by using iterated jet
bundles (the jet bundle of the jet bundle) and the corresponding dual
objects rather than analyzing directly higher jet bundles and hyper-stresses.

Nevertheless, no relation between variational hyper-stresses and the
proposed traction hyper-stresses has be given above. We hope to study
this relation in a forthcoming work.

\begin{acknowledgement*}
Both authors are grateful to BIRS for sponsoring the Banff Workshop
on Material Evolution, June 11-18, 2017, which lad to this collaboration.
R.S.'s work has been partially supported by H. Greenhill Chair for
Theoretical and Applied Mechanics and the Pearlstone Center for Aeronautical
Engineering Studies at Ben-Gurion University.
\end{acknowledgement*}

\noindent %

\end{document}